\documentclass[conference]{IEEEtran}
\IEEEoverridecommandlockouts
\usepackage{cite}
\usepackage{amsmath,amssymb,amsfonts}
\usepackage{algorithmic}
\usepackage{graphicx}
\usepackage{textcomp}
\usepackage{xcolor}
\usepackage{siunitx}
\usepackage{glossaries}
\usepackage{lipsum}
\usepackage{booktabs}
\usepackage{hyperref}
\usepackage{cleveref}
\usepackage{multirow}
\usepackage{tikz}
\usepackage{pgfplots}
\usepackage{pgfplotstable}
\usepackage{threeparttable}

\pgfplotsset{every tick label/.append style={font=\footnotesize}}
\pgfplotsset{every axis label/.append style={font=\small}}

\definecolor{PULPred}{HTML}{A8322C}
\definecolor{PULPblue}{HTML}{1269B0}
\definecolor{PULPgreen}{HTML}{168638}
\definecolor{PULPorange}{HTML}{F29545}
\definecolor{PULPpurple}{HTML}{910569}

\newacronym{soc}{SoC}{system-on-chip}
\newacronym{isa}{ISA}{instruction set architecture}
\newacronym{pulp}{PULP}{parallel ultra-low power}
\newacronym{odrg}{ODRG}{on-demand redundancy grouping}
\newacronym{tcls}{TCLS}{triple-core lockstep}
\newacronym{ecc}{ECC}{error correction code}
\newacronym{wdt}{WDT}{watchdog timer}
\newacronym{seu}{SEU}{single event upset}
\newacronym{set}{SET}{single event transient}
\newacronym{tid}{TID}{total ionizing dose}
\newacronym{rhbd}{RHBD}{radiation hardening by design}
\newacronym{ge}{GE}{gate equivalent}
\newacronym{udma}{$\mu$DMA}{I/O DMA}
\newacronym{gpio}{GPIO}{general purpose input/output}
\newacronym{tcdm}{TCDM}{tightly coupled data memory}
\newacronym{pdk}{PDK}{process desing kit}

\renewcommand{\baselinestretch}{0.99}

\newcommand{\tsmc}{TSMC 28nm}

\DeclareUnicodeCharacter{03BC}{$\mu$}

\def\BibTeX{{\rm B\kern-.05em{\sc i\kern-.025em b}\kern-.08em
    T\kern-.1667em\lower.7ex\hbox{E}\kern-.125emX}}
\begin{document}
\bstctlcite{IEEEexample:BSTcontrol}

\title{
Trikarenos: A Fault-Tolerant RISC-V-based Microcontroller for CubeSats in 28nm
}

\author{\IEEEauthorblockN{Michael Rogenmoser}
\IEEEauthorblockA{\textit{ETH Zurich}\\
Zurich, Switzerland \\
michaero@iis.ee.ethz.ch}
\and
\IEEEauthorblockN{Luca Benini}
\IEEEauthorblockA{\textit{ETH Zurich}\\
Zurich, Switzerland \\
\textit{University of Bologna} \\
Bologna, Italy \\
lbenini@iis.ee.ethz.ch}
}

\maketitle

\begin{abstract}
One of the key challenges when operating microcontrollers in harsh environments such as space is radiation-induced \glspl{seu}, which can lead to errors in computation.
Common countermeasures rely on proprietary radiation-hardened technologies, low density technologies, or extensive replication, leading to high costs and low performance and efficiency.
To combat this, we present Trikarenos, a fault-tolerant 32-bit RISC-V microcontroller SoC in an advanced \tsmc{} technology.
Trikarenos alleviates the replication cost by employing a configurable triple-core lockstep configuration, allowing three Ibex cores to execute applications reliably, operating on ECC-protected memory.
If reliability is not needed for a given application, the cores can operate independently in parallel for higher performance and efficiency.
Trikarenos consumes \SI{15.7}{\milli\watt} at \SI{250}{\mega\hertz} executing a fault-tolerant matrix-matrix multiplication, a 21.5x efficiency gain over state-of-the-art, and performance is increased by 2.96x when reliability is not needed for processing, with a 2.36x increase in energy efficiency.
\end{abstract}


\glsresetall

\section{Introduction}

Space is a challenging environment for electronics: Without protection from the atmosphere and limited protection from earth's magnetic field, high levels of radiation can cause errors (\glspl{seu}) in processors and \glspl{soc}. Furthermore, devices in space are inaccessible once deployed and costly to replace, yet they are often deployed in critical roles requiring high availability. This has led to a requirement for reliability and redundancy in \glspl{soc} for space, often satisfied by time-tested designs and \gls{rhbd} technologies, built on old process nodes and requiring significant power~\cite{hillman_space_2003}. However, as launch costs are decreasing and access to space is becoming more affordable, many smaller satellites and CubeSats are being designed and launched~\cite{akyildiz_internet_2019}. These cannot afford the large power budget required by typical reliable \glspl{soc} for space, with few watts available to the full system.

The open-source RISC-V \gls{isa} has become a key component to enable a new generation of performant and efficient \glspl{soc} for space~\cite{di_mascio_open-source_2021}. The RISC-V \gls{isa} provides the ideal foundation for specialized designs with architectural modifications required for the reliability of \glspl{soc} in space, designed with modularity and extensibility in mind. Thanks to the open-source community and the increasing industry adoption, up-to-date software and compiler support is more readily available for RISC-V-based designs than for designs based on legacy open-source \glspl{isa}, such as PowerPC and SPARC. Furthermore, silicon-proven designs are available open-source, such as the \gls{pulp} platform's PULPissimo~\cite{schiavone_quentin_2018} and can be freely modified and tuned to the requirements of space applications. 

To achieve the increasing computational requirements within the reduced power budgets in CubeSats and nano satellites, denser technologies for on-board processors are required. While commercial technologies have been agressively scaled to improve performance and power efficiency, \glspl{soc} for space lag behind with their scaling, relying on older, larger technology nodes~\cite{hillman_space_2003} and custom \acrfull{rhbd} techniques. Scaling \gls{rhbd} standard cells and tuning \glspl{pdk} of smaller technology nodes is often too costly and requires too much time for the limited quantity of these designs. However, some of these smaller technologies, such as \tsmc{}, have shown tolerance to the destructive effects of radiation ~\cite{borghello_total_2023}, still requiring tolerance to \glspl{seu} both within sequential cells and due to transient effects in combinatorial logic~\cite{di_mascio_open-source_2021}. One of the most effective and affordable methods to implement \gls{seu} tolerance in modern designs is through architectural modifications, which allows for integration of redundancy in modern commercial technology nodes, providing high-performance and efficient \glspl{soc}.

This paper introduces Trikarenos, a reliable RISC-V \gls{soc} in \tsmc{} technology based on the PULPissimo~\cite{schiavone_quentin_2018} architecture. It includes three Ibex~\cite{davide_schiavone_slow_2017} processing cores, by default, operating together in a \gls{tcls} mode. Using \gls{odrg}~\cite{rogenmoser_-demand_2022}, these cores can be unlocked to increase performance when reliability is not required. Furthermore, the memory is protected with an efficient \gls{ecc}, and scrubbers are added to correct latent errors.
Trikarenos runs at up to \SI{270}{\mega\hertz} while consuming only \SI{17}{\milli\watt} in the reliable lockstep configuration. In performance mode, the three cores achieve a speedup of $2.96\times$ on a matrix multiplication benchmark over the \gls{tcls} mode when reliability is not required.

\section{\gls{soc} Architecture}

\begin{figure}[t]
    \centering
    \includegraphics[width=\columnwidth]{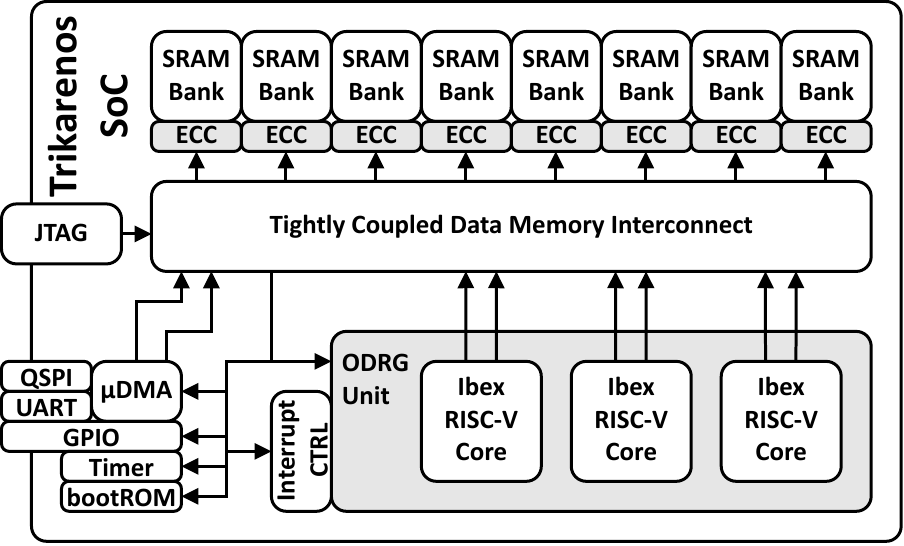}
    \caption{Block Diagram of the Trikarenos \gls{soc}, with the main additions for fault tolerance highlighted.}
    \label{fig:block_diagram}
\end{figure}

The Trikarenos \gls{soc} is shown in \cref{fig:block_diagram} and is based on the PULPissimo~\cite{schiavone_quentin_2018} architecture. As the main processing unit, Trikarenos uses three Ibex~\cite{davide_schiavone_slow_2017} cores featuring the 32-bit RISC-V IMC \gls{isa}. With a 2-stage pipeline and independent instruction and data memory interfaces, the lightweight core is designed for high area efficiency, while providing reasonable performance for microcontrollers. For reliability, the operation of the three cores is locked together with \gls{odrg}~\cite{rogenmoser_-demand_2022} into a \gls{tcls} configuration. In this default \textit{soft-error tolerant} mode, the three cores all receive identical inputs, and their outputs are majority-voted, thereby emulating a single, reliable core to the programmer and the system. If an error is detected, the majority voter immediately ensures it is corrected before affecting the remaining system. To resolve any latent errors within the state of the erroneous core, a re-synchronization routine is executed, storing the full state of the cores in memory on the main software stack, resetting the core, and reloading the state from the stack. If reliability is not needed, \gls{odrg} can be switched into a \textit{performance} mode using control registers within the \gls{odrg} unit wrapping the Ibex cores. This allows all cores to operate independently in parallel, each with its own dedicated inputs and memory interfaces.

The Ibex core's independent instruction and data memory interfaces are connected to a \gls{tcdm} interconnect, giving them access to eight word-interleaved SRAM banks for a total of \SI{256}{KiB}. This allows for fast parallel access to data and instructions for all cores, even in the \textit{performance} mode.
For efficient data protection, each memory bank implements its own \gls{ecc} encoder and decoder, using the Hsiao code~\cite{hsiao_class_1970}, extending the 32-bit data word to 39 bits for storage. With the seven additional bits, the Hsiao code allows for single error correction and double error detection within a data word. The encoder and decoder are designed to encode and correct errors in the same cycle to avoid additional latency. When writing less than a full word, the remaining bits in memory are required to calculate the \gls{ecc} bits correctly; thus, the encoder implements additional logic to load and store the data with an additional cycle, immediately responding to the requester to avoid any unnecessary stalls. To ensure that latent errors within the memory banks are corrected, a memory scrubber continually reads each address and corrects any correctable errors, deferring to external accesses to avoid stalling the system.

The Trikarenos \gls{soc} also includes additional infrastructure and peripherals, also connected to the \gls{tcdm} interconnect.
A boot ROM contains the initial startup code and recovery code to ensure proper recovery for \gls{odrg} in case a fault occurs. Furthermore, an I/O DMA interfaces the \gls{soc} with the outside world, connecting a UART and Quad-SPI peripheral to directly interface with the system interconnect, allowing for independent access to the SRAM banks. The \gls{soc} pads can independently be multiplexed between the different peripherals, and \glspl{gpio}. Finally, a JTAG debug unit allows for programming and debugging of Trikarenos.

\begin{figure}[t]
    \centering
    \includegraphics[width=0.7\columnwidth]{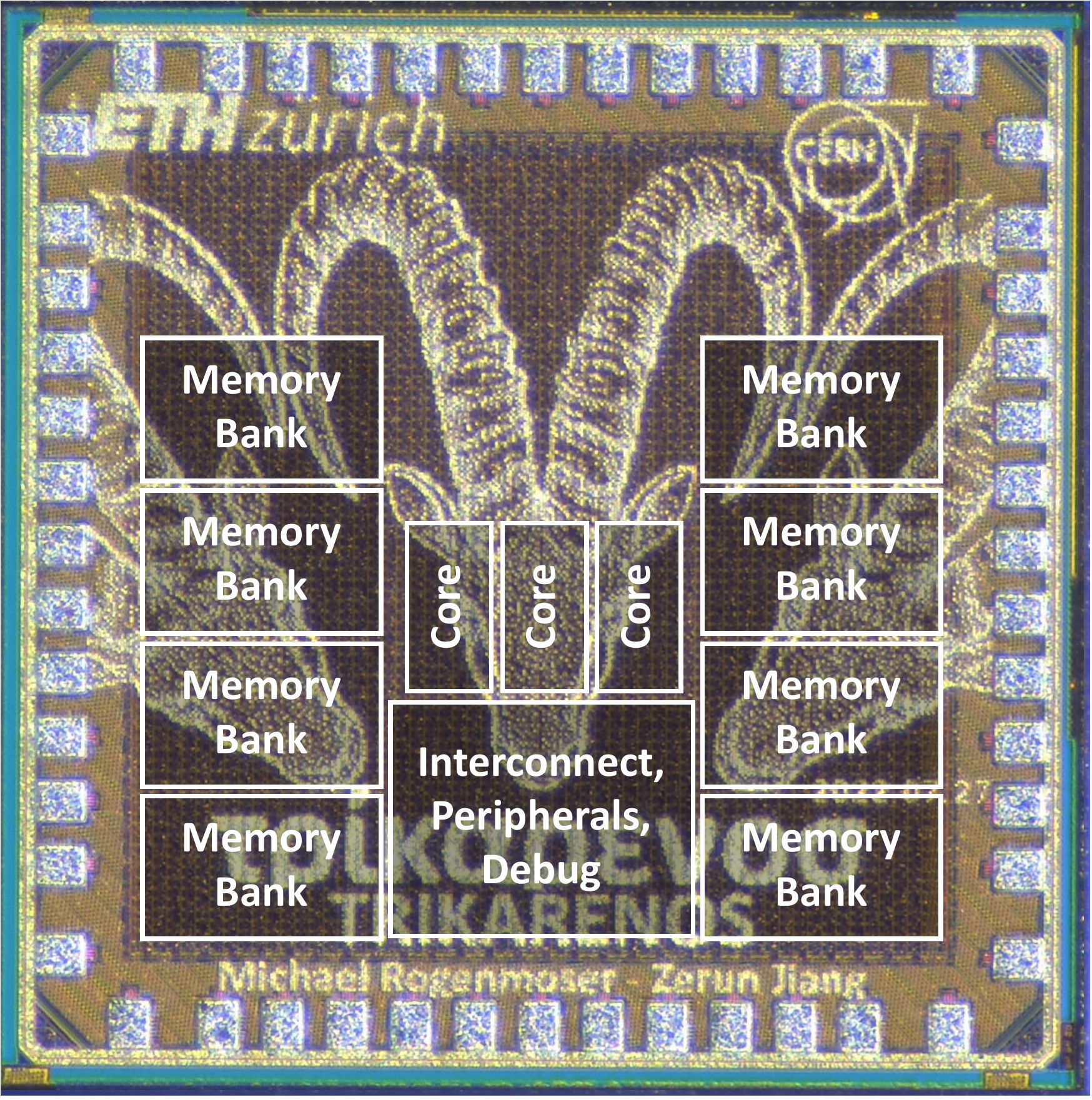}
    \caption{Trikarenos die image with overlayed component highlighting.}
    \label{fig:die_shot}
\end{figure}

\begin{table}[t]
    \centering
    \caption{Physical Characteristics of the Trikarenos \gls{soc} in \tsmc{}}
    \begin{tabular}{@{}cccc@{}}\toprule
        \textbf{Chip Area} & \textbf{SRAM} & \textbf{Core Voltage} & \textbf{Target Frequency} \\
        \SI{2}{\milli\meter\squared} & \SI{256}{KiB} & \SI{0.9}{\volt} & \SI{250}{\mega\hertz} \\ \midrule
        \textbf{Logic Area} & \textbf{SRAM Area} & \textbf{I/O Voltage} & \textbf{Max. Frequency}\\
        \SI{0.15}{\milli\meter\squared} & \SI{0.55}{\milli\meter\squared} & \SI{1.8}{\volt} & \SI{270}{\mega\hertz} \\\bottomrule
    \end{tabular}
    \label{tab:my_label}
\end{table}

\subsection{Design for Testing}

To ensure all designed features can appropriately be tested, Trikarenos adds dedicated testing functionality. All error detectors incorporate dedicated counters and selectable output signals to verify internal behavior and keep track of the errors that occur. Furthermore, the \gls{ecc}-protected memory has custom configuration registers to disable the writing of individual bits, allowing for the software-based injection of errors into select memory banks. Finally, for general testability, the \gls{soc} features scan chains throughout. However, the scan chains were designed such that each core has its own dedicated scan chain. The scan chains can be used to read the complete state information of each core and replace the information with a possibly faulty state. This allows for reliability testing of the \gls{odrg} in a controlled environment.

\section{Results}

\subsection{Implementation}

The \SI{2}{\milli\meter\squared} Trikarenos \gls{soc}, shown in \cref{fig:die_shot}, is implemented in \tsmc{} with a target frequency of \SI{250}{\mega\hertz}. The PULPissimo design occupies an effective core area of \SI{0.7}{\milli\meter\squared} (\SI{2.2}{\mega GE}), sharing the die and I/O with another design. For the implementation, special care was taken to physically separate the three Ibex cores from each other and the remaining logic, ensuring a \SI{20}{\micro\meter} gap without logic around each core. This ensures that any single particle strike cannot simultaneously affect similar elements in separate cores.

\subsection{Performance}

To investigate the performance of the \gls{soc}, a $24\times 24$ matrix-matrix multiplication was implemented and executed on the Trikarenos' Ibex cores, using performance counters within Ibex to determine the cycle count. The chip was tested using the Advantest SoC hp93000 integrated circuit testing equipment to set the frequency and voltage and measure power and energy efficiency, using the application binary to generate patterns for loading with JTAG.

To set up the different processing modes of the three cores, \gls{odrg}'s configuration registers were used, configuring to run in \textit{soft-error tolerant} or \textit{performance} mode through the loaded software. To disable the inactive cores and measure the system's idle power, the cores were instructed to execute the \texttt{wfi} instruction, bringing them into wait-for-interrupt.

\begin{figure}[t]
    \centering
    \pgfplotstableread{
    Voltage Locked Parallel Single Idle Frequency
1.00	26.27	32.55	14.06	8.86    330
0.95	21.20	26.08	12.36	7.01    300
0.90	16.75	20.85	10.77	5.52    270
0.85	12.93	16.32	8.39	4.28    238
0.80	9.62	12.16	6.25	3.20    200
0.75	6.89	8.69	4.49	2.32    162
0.70	4.70	5.92	3.08	1.63    125
0.65	3.00	3.75	2.00	1.10     90
0.60	1.76	2.18	1.20	0.72     59
    }\newpowerdata
    \begin{tikzpicture}
        \begin{axis}[
            width=0.94\columnwidth,
            height=6cm,
            axis y line* = left, 
            xmin=0.58,
            xmax=1.02,
            ymin=0,
            ymax=37,
            xlabel={Core Voltage [\si{\volt}]},
            ylabel={Core Power [\si{\milli\watt}]},
            legend style={nodes={scale=0.8, transform shape}}, 
            legend style={at={(axis cs:0.595, 35)},anchor=north west,},
            y label style={at={(axis description cs:0.12,.5)},rotate=0,anchor=south}
        ]
            \addlegendimage{/pgfplots/refstyle=multiplot:freq}
            \addlegendentry{Max. Frequency}
            \addplot[
                color=PULPgreen,
                mark=square,
                very thick
            ] table [x=Voltage, y=Parallel] {\newpowerdata};
            \addlegendentry{Parallel Cores}
            \addplot[
                color=PULPred,
                mark=square,
                very thick
            ] table [x=Voltage, y=Locked] {\newpowerdata};
            \addlegendentry{Locked Cores}
            \addplot[
                color=PULPorange,
                mark=square,
                very thick
            ] table [x=Voltage, y=Single] {\newpowerdata};
            \addlegendentry{Single Core}
            \addplot[
                color=PULPblue,
                mark=square,
                very thick
                ]
                table [x=Voltage, y=Idle] {\newpowerdata};
            \addlegendentry{Idle}
        \end{axis}
        \pgfplotsset{every axis y label/.append style={rotate=180,yshift=8.8cm}}
        \begin{axis}[
            width=0.94\columnwidth,
            height=6cm,
            xmin=0.58, xmax=1.02,
            ymin=0, ymax=350,
            hide x axis,
            axis y line*=right,
            ylabel={Frequency [MHz]}
        ]
        \addplot[
                color=PULPpurple,
                mark=*,
                very thick
                ]
                table [x=Voltage, y=Frequency] {\newpowerdata};
            \label{multiplot:freq}
      \end{axis}
    \end{tikzpicture}
    \caption{Power measurement results of a 24x24 32-bit Matrix multiplication for different \gls{odrg} configurations at maximum frequency for varying voltage.}
    \label{fig:matmul_pwr}
\end{figure}
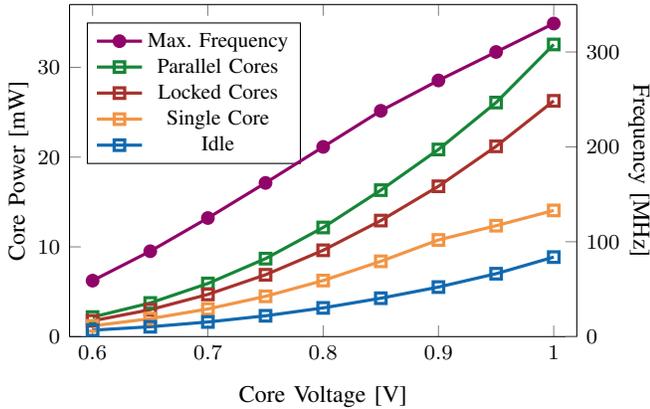

In the default configuration, where all cores are locked together for redundancy, the \gls{soc} consumes \SI{15.7}{\milli\watt} at the designed Frequency of \SI{250}{\mega\hertz} and nominal voltage of \SI{0.9}{\volt}. \Cref{fig:matmul_pwr} shows the maximum achievable frequency and power consumption at that operating point when varying the core voltage, reaching \SI{270}{\mega\hertz} at nominal \SI{0.9}{\volt}. Enabling the \textit{performance} mode of the \gls{soc}, but still relying only on a single core to execute the matrix multiplication, the power consumption drops to \SI{10.3}{\milli\watt}. Comparing the active power to the idle power of \SI{4.9}{\milli\watt} and leakage power of \SI{2.4}{\milli\watt}, where all cores are inactive, the locked configuration has a $2.8\times$ larger power overhead, in line with the $3\times$ more active cores required for the fault-tolerant operation. For both these modes, with only a single core performing the calculation, the matrix multiplication is executed within \SI{187337}{cycles}, at \SI{250}{\mega\hertz} requiring only \SI{0.75}{\milli\second}.

In \textit{performance} mode, with 3 cores available, the $24\times 24$ matrix multiplication can be optimally parallelized across all cores, splitting the workload of the outermost loop across the three cores. Counting the required cycles, the matrix multiplication requires \SI{63130}{cycles}, which amounts to \SI{0.25}{\milli\second} at \SI{250}{\mega\hertz} for a $2.96\times$ speedup over the single-core execution. The faster data processing is also apparent in the power consumption, requiring \SI{19.7}{\milli\watt} for parallel execution, mostly due to more and different data being processed in parallel, leading to higher interconnect and memory activity.

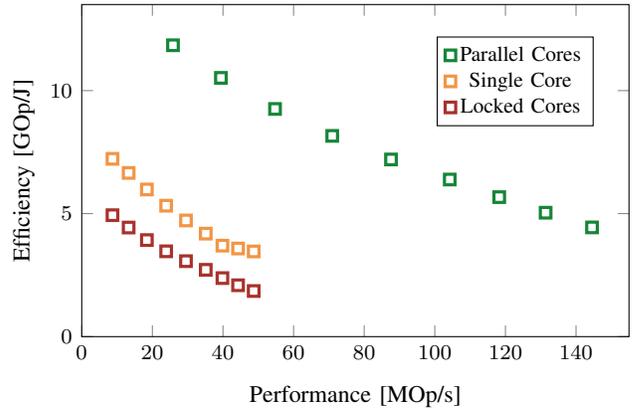
\begin{figure}[t]
    \centering
\pgfplotstableread{
voltage	frequency	locked	parallel	single	singleMOPS	parallelMOPS
1	330	1.853932965	4.440080851	3.463681032	48.70281898	144.5246317
0.95	300	2.088530961	5.037189794	3.582563487	44.27528999	131.3860288
0.9	270	2.379112842	5.670278746	3.698854635	39.84776099	118.2474259
0.85	238	2.716156434	6.386155621	4.18763721	35.12506339	104.2329162
0.8	200	3.067770432	7.201760005	4.722395365	29.51685999	87.59068589
0.75	162	3.468793122	8.160622909	5.321904639	23.90865659	70.94845557
0.7	125	3.924696839	9.255144324	5.981466018	18.44803749	54.74417868
0.65	90	4.434624398	10.51859594	6.654102645	13.282587	39.41580865
0.6	59	4.936209579	11.84417507	7.227318806	8.707473697	25.83925234
}\efficiencydata

    \begin{tikzpicture}
        \begin{axis}[
            width=\columnwidth,
            height=6cm,
            xmin=0,
            xmax=155,
            ymin=0,
            ymax=13.5,
            xlabel={Performance [\si{\mega Op\per\second}]},
            ylabel={Efficiency [\si{\giga Op\per\joule}]},
            legend style={nodes={scale=0.8, transform shape}},
            legend style={at={(axis cs:145, 12.2)},anchor=north east},
            y label style={at={(axis description cs:0.1,.5)},rotate=0,anchor=south}
        ]
            \addplot[
                color=PULPgreen,
                mark=square,
                only marks,
                very thick
            ] table [x=parallelMOPS, y=parallel] {\efficiencydata};
            \addlegendentry{Parallel Cores}
            \addplot[
                color=PULPorange,
                mark=square,
                only marks,
                very thick
            ] table [x=singleMOPS, y=single] {\efficiencydata};
            \addlegendentry{Single Core}
            \addplot[
                color=PULPred,
                mark=square,
                only marks,
                very thick
            ] table [x=singleMOPS, y=locked] {\efficiencydata};
            \addlegendentry{Locked Cores}
        \end{axis}
    \end{tikzpicture}
    \caption{Processing efficiency of a 24x24 32-bit matrix multiplication in different \gls{odrg} configurations at varying voltages, vs. processing performance.}
    \label{fig:efficiency}
\end{figure}

While \cref{fig:efficiency} shows the operating points with maximum frequency at the given voltage, even at the design operating point (\SI{0.9}{\volt}, \SI{250}{\mega\hertz}), \gls{odrg} in Trikarenos allows for a significant efficiency gain. Focusing on energy efficiency, the \textit{soft-error tolerant} mode allows for \SI{2.4}{\giga Op\per\joule}, the performance mode achieves \SI{3.6}{\giga Op\per\joule} with two of the three cores disabled, and \SI{5.6}{\giga Op\per\joule} for parallel computation of the matrix multiplication, an efficiency gain of $1.52\times$ to $2.36\times$. 

\section{Related Work}

\begin{table*}[t]
    \centering
    \caption{Characteristics and performance of state-of-the-art soft-error tolerant \glspl{soc}, comparing with nominal Trikarenos.}
    \begin{threeparttable}
    \begin{tabular}{@{}lcccrrrrrr@{}}\toprule
        \gls{soc} & Class & Cores & \gls{isa} & \multicolumn{1}{c}{Technology} & Performance\tnote{ 1} & \multicolumn{1}{c}{Voltage} & Frequency & \multicolumn{1}{c}{Power} & \multicolumn{1}{c}{Efficiency} \\\midrule
        RAD750~\cite{rea_powerpc_2005} & MCU & 1 & PowerPC & \SI{0.250}{\micro\meter} & \SI{240}{DMIPS} & \SI{2.5}{\volt} & \SI{133}{\mega\hertz} & \SI{5}{\watt}\tnote{ 2} & \\
        RAD55xx~\cite{berger_quad-core_2015} & App. & 1-4 & PowerPC & 45nm & \SI{1398}{DMIPS} & \SI{0.95}{\volt} & \SI{466}{\mega\hertz} & \SI{11.5}{\watt}\tnote{ 2} & \\
        UT700 & MCU & 1 & SPARC & 130nm & \SI{199}{DMIPS} & \SI{1.2}{\volt} & \SI{166}{\mega\hertz} & \SI{4}{\watt}\tnote{ 2} &  \\
        GR716~\cite{johansson_rad-hard_2016} & MCU & 1 & SPARC & UMC 180nm & \SI{90}{DMIPS} & \SI{1.8}{\volt} & \SI{50}{\mega\hertz} & $<$\SI{400}{\milli\watt} & \SI{0.225}{DMIPS\per\milli\watt} \\
        DPC~\cite{van_humbeeck_digital_2016} & MCU & 1 & MSP430 & UMC 180nm &  & \SI{1.8}{\volt} & \SI{120}{\mega\hertz} &  & \\
        AURIX Tricore~\cite{infineon_aurixtricore_2016} & MCU & 1(3) & Tricore & 90nm/65nm & & & \SI{300}{\mega\hertz} &  &  \\
        ARM TCLS~\cite{iturbe_arm_2019} & App. & 1(3) & Armv7-R & ST 65nm & \SI{520}{DMIPS} & \SI{1.2}{\volt} & \SI{311}{\mega\hertz} & \SI{781.9}{\milli\watt} & \SI{0.665}{DMIPS\per\milli\watt} \\
        NOEL-V SoC~\cite{andersson_gaisler_2022} & App. & 1 & RISC-V & ST 28nm & \SI{2256}{DMIPS} & \SI{1.0}{\volt} & \SI{800}{\mega\hertz} &  & \\
        \multirow{2}{*}{\textbf{Trikarenos}} & \multirow{2}{*}{\textbf{MCU}} & \textbf{1(3)} & \multirow{2}{*}{\textbf{RISC-V}} & \multirow{2}{*}{\textbf{\tsmc{}}} & \textbf{\SI{225}{DMIPS}} & \multirow{2}{*}{\textbf{\SI{0.9}{\volt}}} & \multirow{2}{*}{\textbf{\SI{250}{\mega\hertz}}} & \textbf{\SI{15.7}{\milli\watt}} & \textbf{\SI{14.3}{DMIPS\per\milli\watt}} \\
         & & \textbf{3} & & & \textit{\textbf{\SI{666}{DMIPS}}}\tnote{ 3} & & & \textbf{\SI{19.7}{\milli\watt}} & \textit{\textbf{\SI{33.8}{DMIPS\per\milli\watt}}}\tnote{ 3}\\\bottomrule
    \end{tabular}
    \begin{tablenotes}
        \item[1] Performance indicated with Dhrystone MIPS at the operating frequency.
        \item[2] Maximum power dissipation of the full \gls{soc} from the datasheet listed, core power is assumed to be lower thus efficiency was not compared.
        \item[3] Estimated multicore performance based on single core performance and matrix multiplication benchmark speedup.
    \end{tablenotes}
    \end{threeparttable}
    \label{tab:rel_work}
\end{table*}

\Cref{tab:rel_work} shows \glspl{soc} designed for reliability and in most cases, used in space. Due to stringent requirements for reliability, these often are based on older technologies and go through years of testing, and therefore consume significantly more power than commercial counterparts by the time they come to market. Single-core processors for space, such as the US-designed RAD750~\cite{rea_powerpc_2005} or the single-core variants of the RAD55xx~\cite{berger_quad-core_2015} lineup, are built on radiation-hardened process nodes to ensure reliability, at a high cost in area and power consumption. European-designed single-core processors for space, such as the GR716~\cite{johansson_rad-hard_2016}, UT700, and DPC~\cite{van_humbeeck_digital_2016}, are also built on radiation-hardened technologies, frequently augmented with \gls{ecc}. Comparing the performance of the GR716 \gls{soc} to Trikarenos in its reliable configuration, both aimed at reliable, low-power control tasks, we see a $2.5\times$ increase in performance using the Dhrystone benchmark and a $63\times$ increase in efficiency when comparing the upper bound in power consumption of GR716 at \SI{400}{\milli\watt}.

Leveraging lockstepped cores is a common reliability method for automotive designs~\cite{infineon_aurixtricore_2016}, and has been investigated more closely for ARM's Cortex-R5 processor with ARM TCLS~\cite{iturbe_arm_2019}. Here, comparing the implementations using the Dhrystone benchmark, Trikarenos is shown to be $21.5\times$ more energy-efficient while operating in reliable mode. When reliability is not required, Trikarenos can operate in \textit{performance} mode, outperforming the ARM core and offering a $50\times$ improvement in energy efficiency.

One of the most recent designs for space, a \gls{soc} built around the NOEL-V processor~\cite{andersson_gaisler_2022}, is implemented in a \SI{28}{\nano\meter} technology optimized for space applications, but the authors do not report power nor efficiency.

\section{Conclusion}
The paper presents Trikarenos, a fault-tolerant microcontroller \gls{soc} for space applications implemented in \tsmc{}. With three Ibex RISC-V cores in lockstep and \SI{256}{KiB} of \gls{ecc}-protected memory, it offers a reliable microcontroller platform with $21.5\times$ higher efficiency than state-of-the-art. If reliability is not required for a specific task, such as in a CubeSat switching between control and data processing tasks, the three cores can be reconfigured at runtime for a $2.96\times$ increase in performance and $2.36\times$ increase in efficiency.

\section*{Acknowledgements}
The authors thank Gianna Paulin, Luka Milanovic, and Zerun Jiang for their valuable contributions to the research project.
We acknowledge support by the EU H2020 Fractal project funded by ECSEL-JU grant agreement \#877056.
We acknowledge support by TRISTAN, which has received funding from the Key Digital Technologies Joint Undertaking (KDT JU) under grant agreement nr. 101095947. The KDT JU receives support from the European Union’s Horizon Europe’s research and innovation programmes.

\bibliographystyle{IEEEtran}
\bibliography{style,references}

\end{document}